\documentclass{elsart}
\usepackage{graphicx}
\usepackage{amssymb}

\def\Tmelt{\ensuremath{\mathrm{T_M}}}
\def\Tspin{\ensuremath{\mathrm{T_S}}}
\def\gammaSV{\ensuremath{\gamma_\mathrm{SV}}}
\def\gammaLV{\ensuremath{\gamma_\mathrm{LV}}}
\def\gammaSL{\ensuremath{\gamma_\mathrm{SL}}}

\begin{document}

\begin{frontmatter}
\title{NaCl nanodroplet on NaCl(100) at the
        melting point}
\author[sissa,democritos]{T. Zykova-Timan\corauthref{corresponding}},
  \corauth[corresponding]{Corresponding author.}
  \ead{tzykova@sissa.it}
\author[sissa,democritos]{U. Tartaglino},
\author[sissa,democritos]{D. Ceresoli},
\author[ictp]{W. Sekkal-Zaoui} and
\author[sissa,democritos,ictp]{E. Tosatti}
\address[sissa]{International School for Advanced Studies (SISSA-ISAS),
  via Beirut 2, 34014 Trieste, Italy}
\address[democritos]{INFM Democritos National Simulation Center, 
  Trieste, Italy}
\address[ictp]{International Center for Theoretical Physics (ICTP),
  Strada Costiera 11, 34014, Trieste, Italy}

\begin{abstract}
The self-wetting properties of ionic crystal surfaces are studied, using 
NaCl(100) as a prototype case. The anomalously large contact angle
measured long ago by Mutaftschiev~\cite{mutaftschiev75,mutaftschiev97} is
well reproduced by realistic molecular dynamics simulations.
Based on these results, and on independent determinations of the
liquid-vapor ($\gammaLV$) and the solid-vapor interface free energy 
($\gammaSV$)~\cite{unpublished}, an estimate of the solid-liquid interface 
free energy ($\gammaSL$) is extracted.
The solid-vapor surface free energy turns out to be small and similar to 
the liquid-vapor one, providing a direct thermodynamic explanation of the 
reduced wetting ability of the ionic melt.
\end{abstract}

\begin{keyword}
Equilibrium thermodynamics and statistical mechanics \sep
Molecular dynamics \sep
Surface energy \sep
Surface stress \sep
Surface melting \sep
Wetting \sep
Alkali halides
\end{keyword}
\end{frontmatter}

\section{Introduction}\label{sec:intro}
A long known unusual property of molten alkali halide salts is their
remarkable inability to wet their own solid surface at the melting point,
giving rise, in particular to a partial wetting angle as large as 48${}^\circ$
on NaCl(100)~\cite{mutaftschiev75,mutaftschiev97}.
While partial self-wetting must generically arise when there is non-melting 
of the solid surface, as has been long known e.g. for metal 
surfaces~\cite{ditolla95}, there are so far no theoretical studies 
of either effects in ionic systems.
In our work~\cite{unpublished} we have investigated the properties of 
solid NaCl(100) near to the melting point in contact with a droplet of 
its own liquid.

\section{Method of calculation}\label{sec:method}
We carried out extensive classical molecular dynamics simulations of a
nanodroplet of liquid NaCl deposited on the solid NaCl surface.
For NaCl, whose molecular model was described by the Born-Mayer-Huggins 
potential as parameterized by Fumi and Tosi~\cite{tosi64,sangster76},
the Coulomb long-range interactions were treated by the standard 3-dimensional 
Ewald method~\cite{allen87}.
The convergence parameters for the Ewald summations were chosen in order to 
converge the real and reciprocal space sums to a given tolerance 
\mbox{(10${}^{-5}$ Hartree per atom)}.

We simulated NaCl bulk systems and slabs with (100) orientation, 
consisting of about 4000--5000 molecular units. 
The chosen time step was 1 fs, and
the typical simulation time was 200 ps. After verifying that 
the bulk thermodynamic 
properties of solid NaCl (melting temperature, lattice expansion, 
volume jump at melting point~\cite{unpublished}) are very 
well described we 
simulated the approach of a NaCl droplet on NaCl(100) surface.

We prepared a 7200 atom crystalline slab, 8 atomic planes thick 
($15 \times 15\times 4 (\mathrm{NaCl})_4$ conventional cubic cells).
The slab was gradually heated up to the bulk melting point, 
keeping two atomic layers rigidly fixed at the bottom of the slab, 
however with a lattice spacing growing with T, in order 
to simulate a thermally expanding semi-infinite solid. 
Separately we prepared a small NaCl cluster consisting of 500 NaCl 
molecular units. 
The size of this cluster was large enough to yield after melting a 
near-spherical drop, and also a well-defined shape when the drop is 
deposited on the surface. 
The cluster was melted by heating above the bulk melting temperature 
($\Tmelt = 1066\pm 20 $\,K in our model in good agreement with the experimental 
value $\ensuremath{\mathrm{T^{exp}_M}}$ = 1074\,K~\cite{unpublished}), 
and then it was 
equilibrated at T=\Tmelt ~for 100 ps, yielding a well-defined spherical 
liquid droplet of radius 18 \AA.

Finally, we placed the drop near the solid NaCl(100) slab surface, 
the lowermost atoms of the drop a distance 4 \AA\ from the surface 
(fig.~\ref{fig:drop}a).
Zero vertical velocity was assigned to the drop center of mass 
but the drop expanded anyway to touch the surface.
As soon as the drop contacted the surface,
attempts at equilibrating the system at the theoretical melting temperature 
(\Tmelt) failed, and the whole solid NaCl slab melted very quickly, which 
is compatible with the `fragile' nature of non-melting of solid 
NaCl(100)~\cite{unpublished}.
At a much lower temperature than \Tmelt\ (1000 K) on the contrary, 
the substrate remained solid, while  
the drop spread slightly only at the beginning but then crystallized, 
forming a nice ``stepped'' pyramid, made up of (100) facets.
We succeeded finally in equilibrating the liquid nanodroplet on the 
solid surface at the intermediate temperature of 1050K, which is only 
slightly below \Tmelt ~of our model. 
This is described in the following section.

\section{Results}\label{results}
\begin{figure}\begin{center}
  \includegraphics[width=14cm]{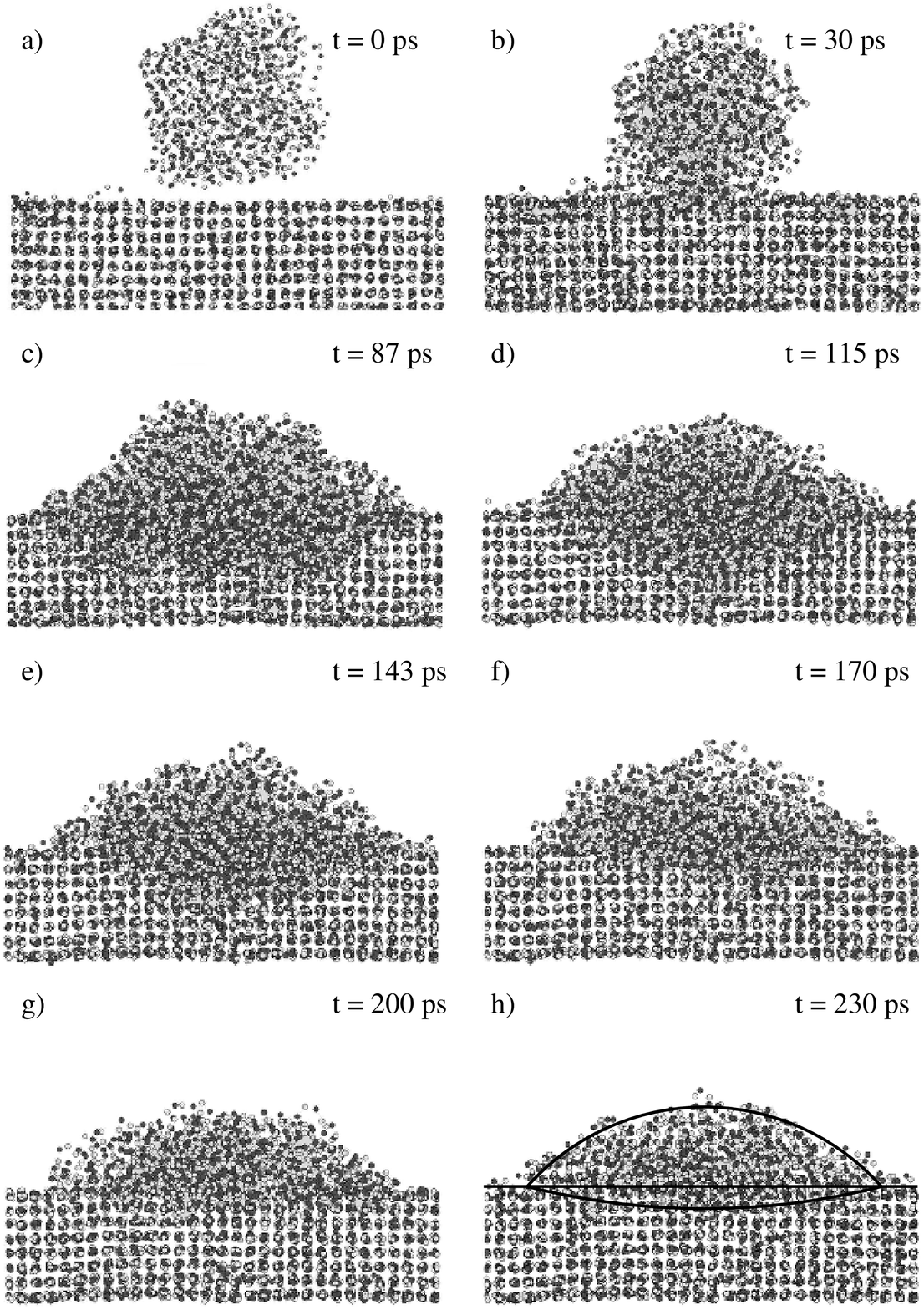}
  \caption{Time evolution of the liquid NaCl drop on NaCl(100).
  Dark and light circles stand for the Na${}^+$ and Cl${}^-$ ions respectively.}
  \label{fig:drop}
\end{center}\end{figure}

\begin{figure}\begin{center}
  \includegraphics[width=8cm]{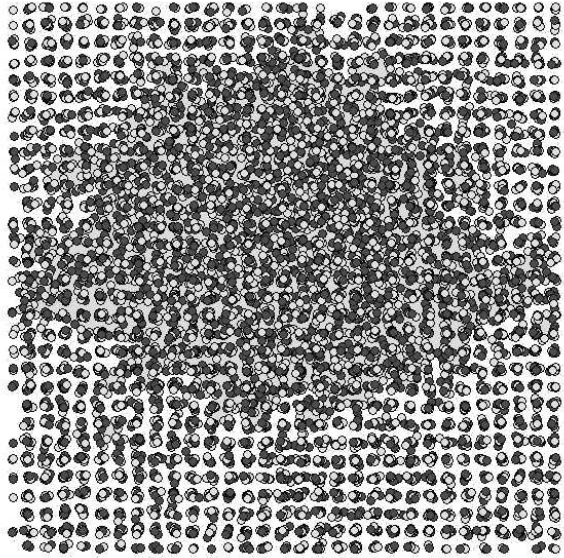}
  \caption{Top view of the liquid droplet on the NaCl(100) after 230 ps.
  Dark and light circles stand for the Na$^{+}$ and Cl$^{-}$ ions respectively.}
  \label{fig:top}
\end{center}\end{figure}

The nanodroplet and the solid slab were separately equilibrated at
1050\,K. During the first 100 ps after contact, the droplet settled
down on the substrate, gradually approaching the final shape
(fig.~\ref{fig:drop}b-d). In the next 130 ps, the droplet survived in a
(not clear if metastable or unstable, but long lived) state without spreading
appreciably (fig.~\ref{fig:drop}e-f). At the end of our simulation, the
droplet -- substrate system looked like fig.~\ref{fig:drop}h. The top
view of fig.~\ref{fig:top} shows that the drop spread almost circularly.

Before proceeding with further descriptions, it is
important to specify the thermodynamics of this situation. Because we
are below \Tmelt\ (even if slightly) the final equilibrium state should consist of a
flat solid NaCl(100) surface, i.e.\ the nanodroplet should completely
spread and recrystallize. That however will take a very long time. While
the nanodroplet exists, it will form an external wetting angle
$\theta_{\mathrm{LV}}$ (fig.\ \ref{fig:drop}), as well as an internal angle
$\theta_{\mathrm{SL}}$. The latter $\theta_{\mathrm{SL}}$ is irrelevant
here, because it depends critically on the temperature and on the time.
In particular at short times and not too far from \Tmelt\ we expect
$\theta_{\mathrm{SL}}\approx 0$. The external angle $\theta_{\mathrm{LV}}$ is
instead significant, as it should equal the macroscopic wetting angle
measured in the bubble experiment\cite{mutaftschiev75,mutaftschiev97}.

\begin{figure}\begin{center}
  \includegraphics[width=12 cm]{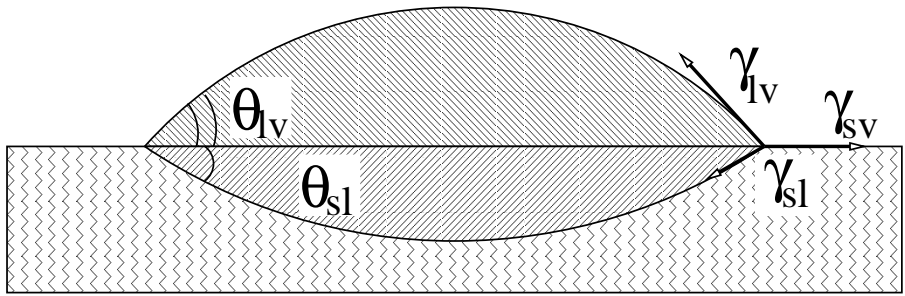}
  \caption{Sketch of the liquid drop partially wetting a solid substrate, 
  showing the balance of the forces acting at the interfaces.}
  \label{fig:schema}
\end{center}\end{figure}

Assuming quasi equilibrium for the liquid nanodroplet in the sense
specified above, this angle will obey Young's equation (fig.\ \ref{fig:schema})
\begin{equation}\label{eq:young}
 \gammaSL + \gammaLV \cos\theta_{\mathrm{LV}} = \gammaSV
\end{equation}
where the $\gamma$'s are the interface free energies.

To determine the external wetting angle $\theta_{\mathrm{LV}}$ of the
nanodroplet we analyzed 100 configurations in last 100 ps. 
The instantaneous atomic positions were plotted in cylindrical
coordinates ($r$ and $z$, where $r$ is parallel to the surface), and from 
the profile of the drop, we determined the best approximation to a portion 
of a sphere, by determining the center position and the radius.
The contact angle follows immediately by simple geometry from these two
quantities. Our best estimate slightly below \Tmelt ~is
$\theta_\mathrm{LV} = 50{}^\circ \pm$ 5${}^\circ$ which is in excellent
agreement with the experiment value at the melting point 
(48${}^\circ$)~\cite{mutaftschiev75}.
At the end of the simulation the internal solid-liquid interface was 
still relatively sharp and flat, consistent with our assumption
$\theta_{\mathrm{SL}}\approx 0$.

The thermodynamical reason leading to the large observed 
$\theta_{\mathrm{LV}}$ can
now be analyzed. As eq.~(\ref{eq:young}) shows, a large value of
$\theta_{\mathrm{LV}}$ may arise either due to a large value \gammaSL, or
to a small \gammaSV, relative to \gammaLV\ or to both.
These free energies at $T\approx\Tmelt$ are all presently unknown. While
simulations are presently under way to calculate
$\gamma_{\mathrm{SV}}$~\cite{unpublished}, we can already anticipate on
purely physical arguments that both factors namely, a large \gammaSL\ and a small
\gammaSV, will indeed occur on NaCl(100). The liquid and the solid
differ enormously, in density and other properties, and this does
suggest that \gammaSL\ has no reason to be small, as it was instead
in the case of
metals. The solid surface free energy is instead small due to a good
cancellation of the Coulomb potential outside the (100) 
surface, and also to a good vibrational free energy 
of the hot solid surface~\cite{unpublished}.

\section{Discussion}\label{discussion}
We can now try to relate further the wetting angle $\theta_{\mathrm{LV}}$
to more microscopic quantities. Di Tolla et al.\ \cite{ditolla95} showed
that for a non-melting metal surface, the angle $\theta_{\mathrm{LV}}$
could be related through a simple model to the surface spinodal
temperature \Tspin, that is the maximum temperature above \Tmelt\ to
which the non-melting surface can be overheated. As discussed in a
companion paper to this~\cite{unpublished}, the very same model
must be improved in order to describe NaCl, owing to the larger 
differences between solid and liquid.
The free
energy gain upon crystallizing the surface at $T=\Tmelt$,
$\Delta\gamma_{\infty}  = ( \gammaLV+\gammaSL ) - \gammaSV$
is related to the partial contact angle through Young's equation (\ref{eq:young}) 
\begin{equation}
 \cos\theta_{\mathrm{LV}} = 1 - \frac{\Delta\gamma_{\infty}}{\gamma_{\mathrm{LV}}}
\end{equation}
If we plug in our calculated value of
$\gammaLV\simeq 104$\,erg/cm$^2$ and $\theta = 50{}^\circ \pm$ 5${}^\circ$,
we obtain $\Delta\gamma_{\infty} = 37$\,erg/cm$^2$.
The large value of $\Delta\gamma_{\infty}$ represents the "binding energy"
of the solid-liquid and liquid-vapor interfaces, to form a single one, 
namely solid-vapor. Assuming a standard form~\cite{pluis}
\begin{equation} 
V(\mathrm{l}) = - \frac{\Delta\gamma_{\infty}}{2}\cos{\frac{2 \pi\mathrm{l}}{a}}\e^{-\mathrm{l}/\xi}
\end{equation}  
for the interface interactions, it leads to a connection~\cite{unpublished}
between $\Delta\gamma_{\infty}$ and the solid surface spinodal temperature \Tspin,
which is the temperature where the solid surface becomes mechanically unstable, 
in the form:
\begin{equation}\label{eq:eq2}
  \Delta\gamma_{\infty}\simeq\rho L a (\Tspin/\Tmelt-1),
\end{equation}   
With our value of $\Delta\gamma_{\infty}$, and $L=4.813\times 10^9\,\mathrm{erg/gr}$, $a=5.9$\,\AA\,
we predict $\Tspin= 1210 \,K$, very close to that seen in simulations.~\cite{unpublished}

\section{Conclusion}\label{conclusion}
In conclusion, molecular dynamics simulations indicate that metastable 
liquid NaCl nanodroplets on solid NaCl(100) can exist for some time
close to \Tmelt. They are found to exhibit
an unusually large partial wetting angle, whose value is in agreement with 
the large macroscopic wetting angle observed in the bubble experiment.
Thermodynamically this appears to be caused by an
unusual small value of the solid-vapor interface free energy, as well 
as by a large solid-liquid interface free energy. 
The free energy differences extracted via Young's equation can also be 
connected well with other properties, such as the solid surface spinodal 
temperature.

\section*{Acknowledgments}\label{acknow}
Project was sponsored by Italian Ministry of University and Research, through
COFIN02, COFIN03, and FIRB RBAU01LX5H; and by INFM, through PRA NANORUB and
and ``Iniziativa Trasversale calcolo parallelo''.
Calculations were performed on the IBM-SP4 at CINECA, Casalecchio (Bologna).


\end{document}